\documentclass[aps,prx,twocolumn,groupedaddress]{revtex4-2}
\usepackage{graphicx}
\usepackage{amssymb}
\usepackage{amsmath}
\usepackage{xcolor}

\begin{document}


\title{Damping of a micro-electromechanical oscillator in turbulent superfluid $^4$He: A novel probe of quantized vorticity in the ultra-low temperature regime}

\author{C. S. Barquist}
\email{cbarquist@ufl.edu}
\author{W. G. Jiang}%
\author{K. Gunther}
\author{N. Eng}
\author{Y. Lee}
\email{ysl@ufl.edu}
\affiliation{Department of Physics, University of Florida, Gainesville, Florida, 32611, USA}%

\author{H. B. Chan}
\affiliation{Department of Physics, Hong Kong University of Science and Technology, Clear Water Bay, Kowloon, Hong Kong}%

\date{\today}
	
\begin{abstract} We report a comprehensive investigation of the effects of quantum turbulence and quantized vorticity in superfluid $^4$He on the motion of a micro-electromechanical systems (MEMS) resonator. We find that the MEMS is uniquely sensitive to quantum turbulence present in the fluid. To generate turbulence in the fluid, a quartz tuning fork (TF) is placed in proximity to the MEMS and driven at large amplitude. We observe that at low velocity, the MEMS is damped by the turbulence, and that above a critical velocity, $v_c \simeq 5\,$mm\,s$^{-1}$, the turbulent damping is greatly reduced. We find that above $v_c$, the damping of the MEMS is reduced further for increasing velocity, indicating a velocity dependent coupling between the surface of the MEMS and the quantized vortices constituting the turbulence. We propose a model of the interaction between vortices in the fluid and the surface of the MEMS. The sensitivity of these devices to a small number of vortices and the almost unlimited customization of MEMS open the door to a more complete understanding of the interaction between quantized vortices and oscillating structures, which in turn provides a new route for the investigation of the dynamics of single vortices.
\end{abstract}


\maketitle

\section{Introduction}
Similar to classical fluids, superfluids, such as $^4$He, $^3$He, and Bose-Einstein condensates of ultra cold gases, can also become turbulent \cite{Donnelly1986,vinen2002quantum,Vinen2006,Barenghi2014,Tsatsos2016}. This phenomenon goes by the name {\it quantum turbulence} (QT). Quantum turbulence differs from classical turbulence in at least two key ways. Firstly, in QT, all of the circulation in the superfluid component is due to quantized vortices. Analogous to vortices in type II superconductors, quantized vortices in a superfluid consist of circulating superfluid around a normal core of diameter $a_0$. The quantum of circulation is $\kappa = h/m$ \cite{Donnelly2005}, where $h$ is Planck's constant and $m$ is the mass of the boson constituting the superfluid.  For $^4$He, the vortex core diameter estimated from the nonlinear Schr{\"o}dinger equation is $a_0 \simeq 10^{-10}$\,m \cite{Donnelly2005} and $\kappa= 9.97\times 10^{-8}$\,m$^{2}$\,s$^{-1}$. Because the energy of a quantized vortex is proportional to the square of the circulation \cite{Donnelly2005}, doubly quantized vortices are unstable, with the fluid preferring two singly quantized vortices instead. Therefore, all of the circulation is due to singly quantized vortices, and QT may be understood as a tangle of these identical vortices. Secondly, the superfluid differs from the classical fluid in that it has zero viscosity, and viscous dissipation is absent.

Despite these differences, in some ways QT is remarkably similar to classical turbulence. In both $^3$He \cite{Bradley2006,Bradley2011} and $^4$He  \cite{Walmsley2007,Walmsley2008,Zmeev2015,Walmsley2017a}, the decay rate of turbulent energy was found to have the same time dependence as predicted by the classical Kolmogorov-Ohbukov theory. In $^4$He, the quintessential $k^{-5/3}$ law for turbulent fluctuations and intermittency has also been observed \cite{Maurer1998}. Remarkably, the similarities between quantum and classical turbulence, do not exist only at high temperatures, where the superfluid and normal fluid are coupled by mutual friction. In this regime, it might be expected that the turbulence in the superfluid inherits classical characteristics through its interaction with the normal fluid. However, even in the ultra-low temperature regime, where the normal fluid is effectively absent, the superfluid is able to mimic classical turbulent flow on scales larger than the average inter vortex distance, $\ell = L^{-1/2}$, where $L$ measures vortex line length per unit volume. By polarizing and forming bundles, quantized vortices are able to generate flow on all scales between $\ell$ and the system size \cite{Baggaley2012a}. In this way, pure superfluid turbulence can behave quasi-classically \cite{Walmsley2014}. Although, at scales similar to and smaller than $\ell$, the individual nature of the quantized vortices becomes apparent and the flow loses any classical character.

In the ultra-low temperature regime, in the absence of viscous damping, how is the energy in quantum turbulence dissipated? It is well established through experiment \cite{Walmsley2007,Walmsley2008,Zmeev2015,Walmsley2017a,Bradley2006,Bradley2011,Skrbek2012} and simulation \cite{tsubota2000dynamics,kobayashi2005kolmogorov,barenghi2008reynolds,Baggaley2012a,Baggaley2012,Sasa2011a} that at scales much larger $\ell$, turbulent energy is transferred to smaller scales in a manner similar to the Richardson cascade \cite{Frisch1995} of classical turbulence. However, at scales similar to $\ell$ this process must stop, and a new process must take over. It is thought that the energy is carried to smaller scales by the Kelvin wave cascade on individual vortices before it is radiated away as phonons. However, direct experimental evidence of this process has not been observed, due to the lack of an appropriate experimental probe.

When the normal fluid is absent, powerful probes, such as second sound and tracer particles, cease to function. There has been significant success with injected electrons \cite{Walmsley2007,Walmsley2008,Zmeev2015,Walmsley2017a}. Currently, this method has only been able to measure average properties about large scale flow, such as the average $L$. In the past two decades, small oscillating objects, such as vibrating wires \cite{Yano2005,Yano2005a,hashimoto2007control,hashimoto2007switching,Yano2007,goto2008turbulence,yano2008instability,yano2009transition,nago2010observation,nago2010vortex,Yano2010,nago2011time,nago2013vortex,kubo2013time,oda2014observations,fisher2001generation,bradleyTurbulence,bradley2005emission,Bradley2005a,Bradley2006,Bradley2008,bradley2009transition2,Bradley2011,bradley2013onset,Fisher2014,bradley2000repetitive}; tuning forks \cite{blavzkova2007quantum,blavzkova2007transition,blavzkova2008vibrating,blavzkova2009generation,bradley2009transition1,bradley2009damping,Garg2012,ahlstrom2014frequency,Bradley2014,bradley2016probing,Schmoranzer2016,Ahlstrom2014,Duda2017,Gritsenko2018}; micro-spheres \cite{jager1995turbulent,jager1995translational,jager1996turbulent,schoepe2010statistics,schoepe2013shedding,schoepe2015breakdown,niemetz2002intermittent}; and vibrating grids \cite{vinen2004nucleation,nichol2004experimental,Nichol2004a,bradley2005emission,charalambous2006experimental,Efimov2009,Bradley2012}, have successfully investigated many properties of QT in $^4$He and $^3$He over a large range of temperatures, including the ultra-low temperature regime. They have been particularly useful in understanding the generation of QT and the crossover from laminar to turbulent flow. However, in $^4$He, none are continuously sensitive to externally applied turbulent flow, as they begin to generate their own turbulent flow after being exposed to vortices from the turbulent flow. This precludes them from being able to measure important quantities, such as the fluctuations in turbulent energy or in line density, $L$.

In this work, we investigate the effects of QT and quantized vorticity on the motion of a  micro-electromechanical systems (MEMS) resonator in the ultra-low temperature limit of $^4$He. Previously, similar MEMS resonators have been used to study superfluid \cite{Zheng2016,Zheng2017,Zheng2017a} and normal fluid \cite{gonzalez2016temperature} $^3$He. To generate QT, a quartz turning fork (TF) is placed in proximity to the MEMS and driven with large amplitude. We find that the MEMS is uniquely sensitive to vortices and is able to continuously monitor the turbulent flow. We also find that the coupling between the MEMS and quantized vortices is velocity dependent, with a critical velocity of $5\,$mm\,s$^{-1}$ separating two distinct regimes of coupling. While these are not observations of the elusive Kelvin wave cascade, the sensitivity of these devices to a small number of vortices and the almost unlimited customization of MEMS open the door to a more complete understanding of the interaction between quantized vortices and oscillating structures, which in turn provides a new route for the investigation of the dynamics of single vortices and turbulent fluctuations in the ultra-low temperature regime.

\section{Experimental}
\subsection{Devices}

\begin{figure*}
	\centering
	\includegraphics[width=0.9\linewidth]{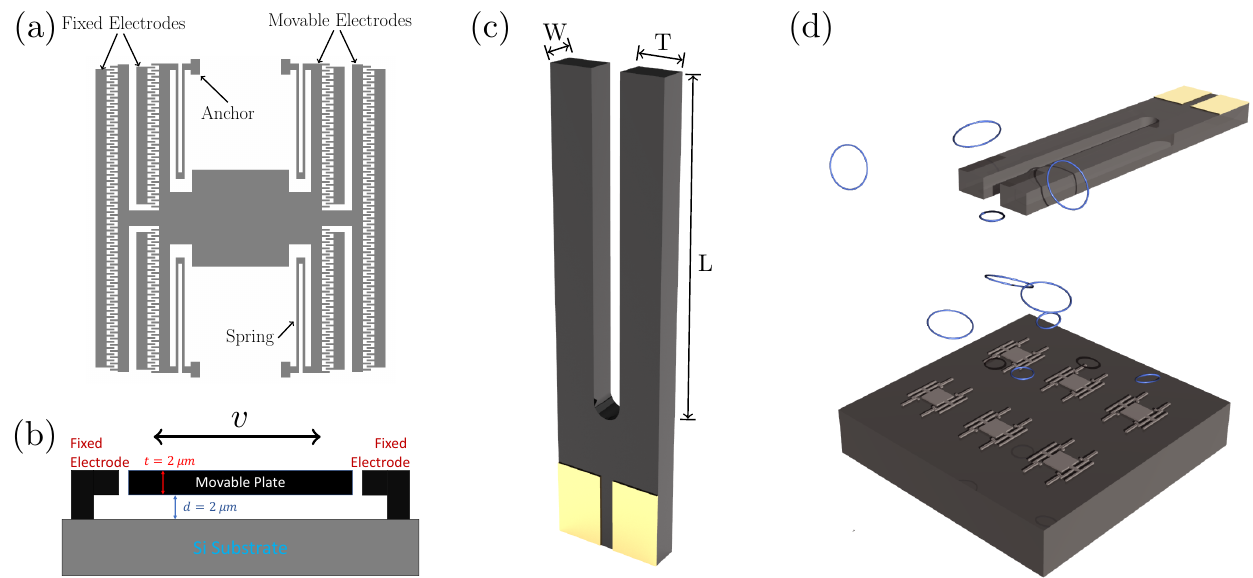}
	\caption{Schematic diagram of the oscillators used in this experiment (a) A scale diagram of the MEMS. The MEMS consists of a $125\times 125 \,\mu$m$^2$ square plate, and is suspended above the substrate by springs which are anchored to the substrate at the anchor points. The device is actuated and detected through the capacitively coupled electrodes on either side. (b) A cartoon of the cross section of the MEMS (not to scale). The MEMS has a uniform 2\,$\mu$m thickness, and is suspended above the substrate by 2\,$\mu$m. The arrows indicate the direction of motion for the primary resonance mode discussed. (c) Cartoon of the TF used to drive turbulence in the experiment. The dimensions of the fork are $W=0.10\,$mm, $T=0.23\,$mm, and $L=2.36\,$mm. The tines oscillate in the plane of the fork and in antiphase to one another.  (d) Schematic of the experimental setup. The TF is located $\simeq 3\,$mm above the MEMS device. When the TF is driven with large amplitude, turbulence is formed and vortex rings are ejected from the tangle, which interact with the MEMS.}
	\label{fig:1}
\end{figure*}

The MEMS device used for this study is 2\,$\mu$m thick and consists of a $125\times 125\,\mu$m$^2$ square plate with two rows of capacitively coupled comb electrodes on two opposite sides of the plate. The device is suspended 2$\,\mu$m above a substrate by four springs, which allows for a fluid film to be formed beneath the device. A diagram of the device is shown in Fig.\,\ref{fig:1}(a) along with a cartoon of the cross section of the device in Fig.\,\ref{fig:1}(b). Due to the geometry of the MEMS device, there are several modes of oscillation, which are illustrated in Refs. \cite{Gonzalez2013,csbthesis}. In this work, we only study the behavior of the shear mode, which has its motion directed in the plane of the device, as shown in Fig.\,\ref{fig:1}(b). In contradistinction to most of the resonators mentioned previously, when oscillating in the shear mode,  the whole device is displaced equally, and the velocity is uniform. This is an advantage of our device, as non-uniform velocity profiles have the tendency to blur the measurements of velocity dependent phenomena.

\begin{figure}
	\centering
	\includegraphics[width=1\linewidth]{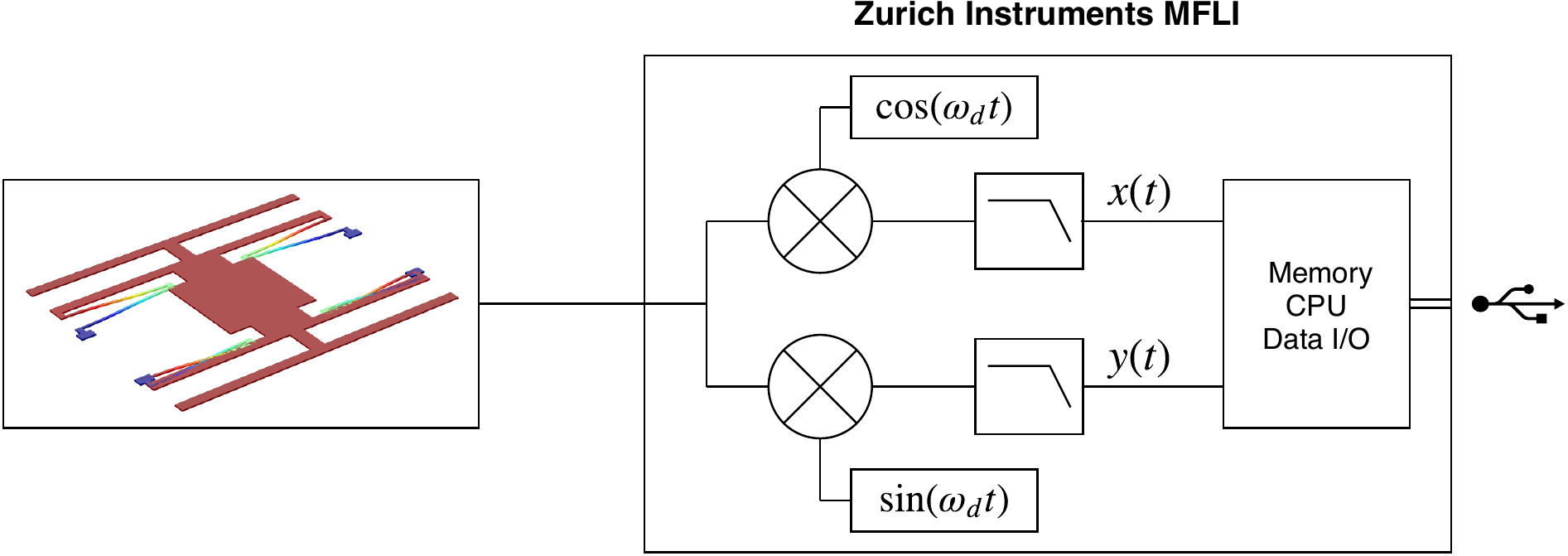}
	\caption{The MEMS and TF are measured through a lock-in referenced at $\omega_d = 2\pi f_d$, where $f_d$ is the driving frequency for each respective device. This setup is employed for both time and frequency domain measurements, and allows for the collection of both phase and amplitude information.}
	\label{fig:x}
\end{figure}

The device is asymmetrically driven and detected via the comb electrodes. An oscillating voltage, $V(t) = V_f \cos(2\pi ft/2)$, is applied to one side of the device, which creates a force due to the gradient of the electrostatic energy, $F = \frac12 V^2\frac{dC}{dx}$, where $C$ is the capacitance between the movable electrode and the fixed electrode, and $x$ is the displacement of the device from its equilibrium position. The driving force is then 
\begin{equation}
F_d = \frac14 \beta V_f^2\cos(2\pi ft),
\end{equation}
where $\beta = \frac{dC}{dx}\rvert_{x=0} = 1.44\,$nF\,m$^{-1}$ (in helium) is called the {\it transduction factor} \cite{csb-2020-prx-supp}. Because the force depends on the square of the voltage, the force is transduced at twice the excitation frequency, and also at $f=0$. The DC component of the force, $\frac14 \beta V_f^2$, negligibly shifts the equilibrium position, and may be ignored.

The displacement is detected by using the electrodes on the other side of the device. A DC bias voltage, $V_b$, applied to the electrodes induces a charge on the electrodes. As the device is displaced the capacitance varies by $\delta C$, which varies the charge by $\delta q = V_b\delta C = V_b \beta x$. The change in the charge is measured by a charge sensitive amplifier with amplification $\alpha = 0.67\,$pF$^{-1}$. The displacement can be calculated as
\begin{equation}
x = \frac{V_o}{\alpha \beta V_b},
\end{equation}
where $V_o = \alpha \delta q$ is the voltage measured at the output of the charge sensitive amplifier.

At the low temperatures used in this work, damping of the device is minimal and the quality factor, $Q$, exceeds 10$^5$. Because of this, the nonlinear nature of the device is apparent, see Fig.\,\ref{fig:3}. The nonlinearity observed in this device is attributed to the nonlinear variation of the capacitance. After a carefully accounting for the nonlinear variation in capacitance  \cite{csb-2020-prx-supp}, it is found that two additional forces need to be considered: a modification to the spring constant, $\frac12 V_b^2c_1x$, and a nonlinear spring restoring force, $\frac12 V_b^2c_3 x^3$, where $c_1$ and $c_3$ are constants that depend on the geometry of the electrodes. Additional forces also arise which are proportional to $V_f^2$; however, $(V_f/V_b)^2 < 0.002$ for all measurements, so they may be ignored. Including these nonlinearities the equation of motion for the MEMS is 
\begin{equation}
\ddot{x} + 2(\Gamma_1 + \Gamma_2 x^2)\dot{x} + \omega_0^2 x + \alpha_3 x^3 = g_0 \cos(2\pi ft), 
\label{eq:eom}
\end{equation}
where $2\Gamma_1 = \Delta \omega$ is the full width at half max (FWHM) of the resonance at low amplitude, $2\Gamma_2$ characterizes the nonlinear damping intrinsically present in the silicon, $\omega_0^2 = (k-\frac12V_b^2c_1)/m$, $k$ is the mechanical spring constant, $m$ is the mass, $\alpha_3 = -\frac12 V_b^2c_3/m$, and $g_0 = \beta V_f^2/4m$.

The TF used in this work is the commercially available Epson C-002RX. These forks are typically used as timing devices for integrated circuits, and they are designed to have a frequency of 2$^{15}=32768$\,Hz. They come packaged in a hermetically sealed vacuum can, which is lathed off exposing the fork. A 3D rendering of the TF is shown in Fig.\,\ref{fig:1}(c). The TF is made of single crystal quartz with metal electrodes patterned on the tines (not shown). The fork consists of two large aspect ratio tines of length $L = 2.36$\,mm, width $W = 0.10$\,mm, and thickness $T = 0.23$\,mm. When oscillating in its fundamental mode (the only mode used in this work) the tines oscillate in anti-phase to one another in the plane of the fork. It is actuated and its motion is detected by taking advantage of the piezoelectirc property of quartz. The electrical properties of the TF are characterized by a single quantity, the {\it fork constant}, $a$ \cite{blaauwgeers2007quartz}.  Similar to the MEMS, the fork is driven by applying an alternating voltage, $V(t) = V_f \cos(2\pi f t)$, directly to one of the electrodes on the fork. This applies a force on the fork proportional to $V(t)$:
\begin{equation}
F_d = \frac12 a V_f\cos(2\pi ft).
\label{eq:fork_force}
\end{equation}
In contradistinction to the MEMS, the force on the TF is at the same frequency as the applied excitation. As the fork oscillates, a current is generated proportional to the velocity of the fork, $v(t)$:
\begin{equation}
I(t) = a v(t).
\label{eq:fork_current}
	\end{equation}
It should be noted that the fork tines do not move with uniform velocity, {\it i.e.,} the tines have the maximum velocity $v$ at the tip and zero velocity at the base. The current is then amplified using a transimpedance amplifier with amplification $\alpha = -10\,$k\,$\Omega$. The output voltage, $V_o = \alpha I$, can then be directly related to the velocity using Eq.\,\ref{eq:fork_current}. Calibration of the fork and determination of the fork constant are discussed further in the supplementary information. For the fork used in this work it was found that $a= 2.84\,\mu$C\,m$^{-1}$ \cite{csb-2020-prx-supp}.

\subsection{Measurement Technique}
To study the behavior of the MEMS in the presence of turbulence, the MEMS and the TF are situated in close proximity, with the TF 3\,mm above the MEMS, as depicted in Fig.\,\ref{fig:1}(d). The devices are located inside of a copper cell with a cylindrical volume of about 2\,cm$^{3}$. The cell is affixed just below the mixing chamber stage of a dilution refrigerator. To generate turbulence, the TF is driven with a large amplitude. For TFs, the transition to turbulent flow and their behavior in the turbulent regime has been studied extensively over the past decade, in both $^4$He and $^3$He-B \cite{blavzkova2007quantum,blavzkova2007transition,blavzkova2008vibrating,blavzkova2009generation,bradley2009transition1,bradley2009damping,Garg2012,ahlstrom2014frequency,Bradley2014,bradley2016probing,Schmoranzer2016,Ahlstrom2014,Duda2017,Gritsenko2018}. Because the MEMS is most sensitive to the effects of vortices when the damping is smallest, all of the measurements made in the turbulent regime are made at the lowest attainable temperature of 14\,mK to avoid excess damping due to the presence of normal fluid.

\begin{figure}
	\centering
	\includegraphics[width=\linewidth]{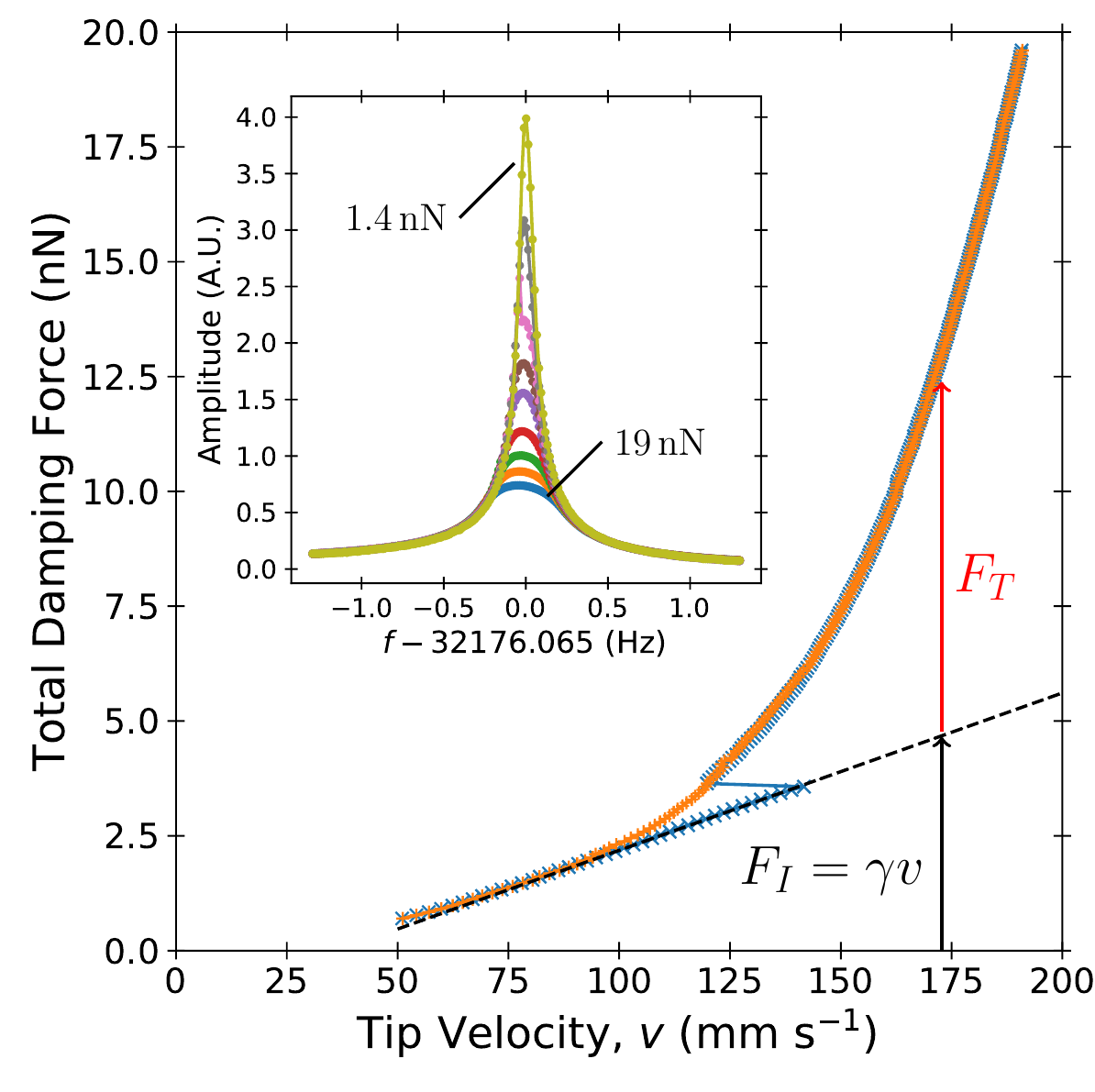}
	\caption{Damping force on resonance for various velocities at 14\,mK in He-II for increasing (blue) and decreasing (orange) velocities. The change of slope of the TF's velocity dependence on force is associated with the generation of turbulence. The dashed line is an extrapolation of the potential flow regime to higher velocities. The excess damping force due to turbulence is calculated by subtracting off the extrapolated damping force from the potential flow regime. Inset: Frequency sweeps of the TF in He-II at 14\,mK for driving forces in the range 1.4\,--\,19\,nN. Here, the responses of the device have been normalized by the driving excitation. The data collapse to a universal curve on the low velocity tails, but fail to collapse on resonance. This indicates that at low velocities the damping is linear and becomes nonlinear beyond $v_c$, due to the generation of turbulence.}
	\label{fig:2}
\end{figure}

To characterize the generation of turbulence by the TF, we measure the velocity of the TF on resonance as the driving force is varied. This allows us to measure the velocity dependent damping force experienced by the TF. When the TF is on resonance, the force is in phase with the velocity, and we may equate the driving force with the damping force. In order to remain on resonance, a feedback loop was employed, taking advantage of the property that the quadrature component of the TF signal passes through zero on resonance. The feedback loop is implemented in a LabVIEW program and adjusts the frequency of excitation until the quadrature component is within some specified distance from zero. Figure\,\ref{fig:2} shows the results of this measurement for increasing and decreasing driving force. At low velocity, the damping force on the TF is roughly proportional to the velocity, $F\propto v$. This is identified as the laminar regime. More accurately, the flow around the TF in this regime is potential, because the viscous normal fluid is absent at this temperature. As the velocity is increased beyond about 140\,mm\,s$^{-1}$, the velocity jumps to a lower value, indicating a sudden increase in the damping, and the damping force is no longer proportional to the velocity. This is identified as the turbulent regime. As the velocity is reduced in the turbulent regime, the damping force continues to follow the new power law until it eventually crosses over into the laminar regime. The velocity where the turbulent and potential regimes merge is identified as the critical velocity, $v_c$. For our TF, $v_c = 90\,$mm\,s$^{-1}$.

This behavior is different from the generation of turbulence in classical fluids. In a classical fluid, there is no critical velocity and the onset of turbulence is continuous. Here, in pure superfluid, there is no turbulent flow or emission of vorticity below $v_c$ \cite{bradley2009transition1,nago2013vortex,jager1995turbulent}. The difference arises because vorticity in superfluid $^4$He is nucleated extrinsically from preexisting remnant vortices pinned to the surfaces of the oscillating structures. The growth of these remnant vortices occurs through the Glaberson-Donnely instability \cite{glaberson1966growth}, which only happens above a critical velocity determined by the size of the largest vortex pinned to the device.

To further illustrate the effects of turbulence generation on the TF, a set of frequency sweeps of the TF are shown in the inset of Fig.\,\ref{fig:2}. There, the sweeps are scaled by the driving excitation. Collapse of the data onto a single universal curve after scaling indicates that the fork is in the linear regime. Here, it is clearly seen that the data on resonance do not overlap, and that the data measured with larger driving force are situated toward the bottom, which indicates increased damping at higher velocities. However, the tails of the resonance still collapse to a single curve, confirming that the excess damping is only present above $v_c$.

The effects of turbulence on the MEMS device was investigated by performing measurements in both the frequency domain and time domain. For both types of measurement, the output of the MEMS, after the preamp, was fed into a lock-in amplifier referenced at the frequency of the driving force, $f_d$ (twice the excitation voltage frequency). This allows us to collect amplitude and phase information by measuring the components of the signal in and out of phase with the driving force, see Fig.\,\ref{fig:x}.

The frequency response of the MEMS is measured by driving the MEMS with a fixed amplitude while varying the driving frequency through the resonance. In the absence of turbulence, the frequency response of the device is modeled by Eq.\,\ref{eq:eom} \cite{Lifshitza,Nayfeh1995}. The displacement of the MEMS has the following form $x(t) = A(\omega)\cos(\omega t + \phi(\omega))$ with
\begin{equation}
A(\omega) = \frac{g_0}{\sqrt{(\omega_0^2-\omega^2 + 2 \Pi A\omega_0)^2 + \omega^2(2\Gamma_1 + \frac12\Gamma_2 A^2)^2}}
\label{eq:frequency_amplitude}
\end{equation}
and
\begin{equation}
\tan(\phi(\omega)) = \frac{-2\omega (\Gamma_1 + \frac12 \Gamma_2 A^2)}{\omega_0^2 - \omega^2 +2\Pi A^2 \omega_0}.
\label{eq:frequency_phase}
\end{equation}

\begin{figure}
	\centering
	\includegraphics[width=0.85\linewidth]{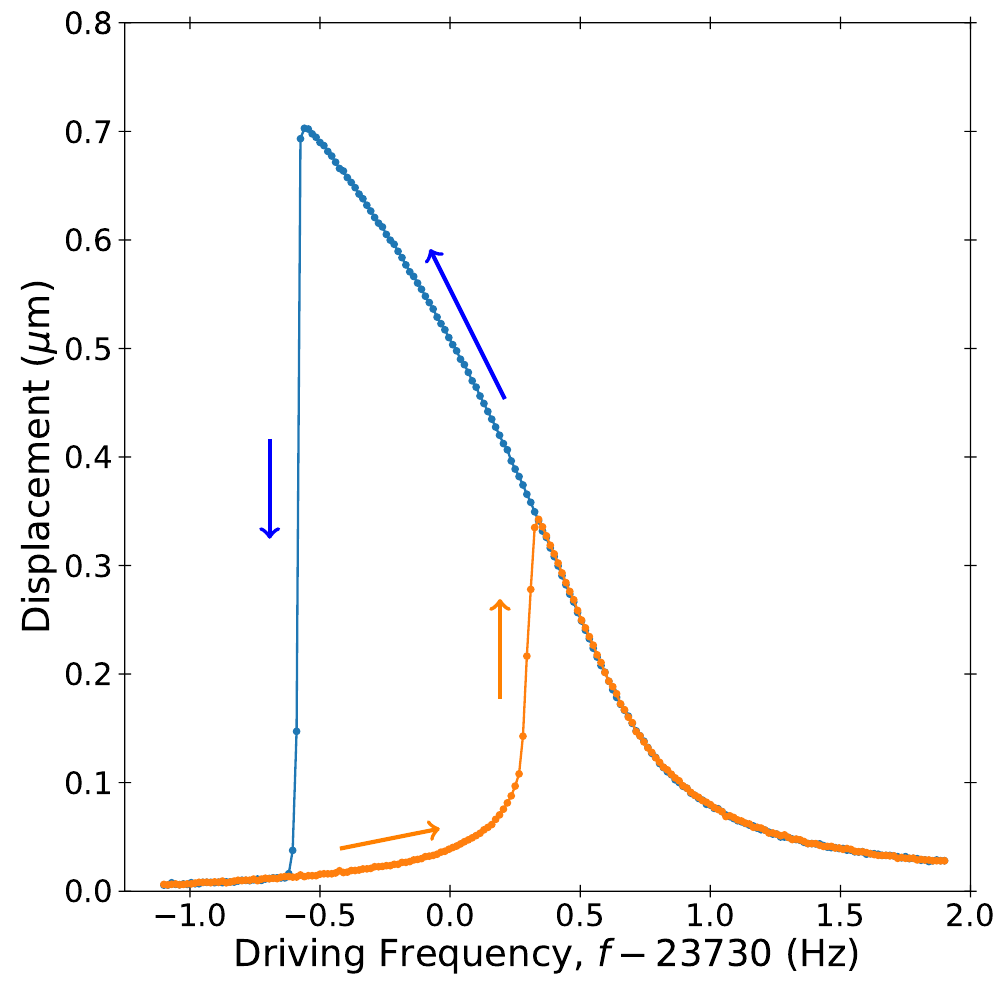}
	\caption{Frequency response of the device in vacuum at 6\,mK, demonstrating the hysteresis between upward and downward frequency sweeps through the resonance.}
	\label{fig:3}
\end{figure}

Here, $\Pi = \frac38 \frac{\alpha_3}{\omega_0}$. Above a critical amplitude, $a_c$, which depends on all of the resonance parameters, a hysteresis appears between sweeping through the resonance with increasing and decreasing frequency \cite{Lifshitza}. Figure\,\ref{fig:3} shows an upward and downward frequency sweep through the resonance of the MEMS, with the directions of the sweeps indicated by the arrows. A clear hysteresis is observed, which is characteristic of the Duffing nonlinearity. The intrinsic ({\it i.e.} not due to the fluid) nonlinear nature of the damping can be seen in Fig.\,\ref{fig:5}(a), where the intrinsic damping measured in vacuum is shown as the solid green curve.

For the measurements that follow, only downward sweeps through the resonance (down sweeps) are considered. This is because we are primarily focused on the damping force experienced by the MEMS. Because our device possess a spring softening nonlinearity, the down sweeps contain the resonance peak (maximum displacement). As is true for the linear resonator, at the peak of the nonlinear resonance, the force and velocity are in phase, {\it i.e.,} $\phi(\omega_{peak}) = \pi/2$. Therefore, by measuring the peak velocity we may equate the driving force with the damping force and map the relationship between MEMS velocity and damping force.

The time domain response of the MEMS is measured by observing its free decay (ringdown). To do this, the MEMS is first energized by driving it at a frequency close to resonance. For these measurements, we wish to start at large amplitude on the upper branch of the resonance (blue curve of Fig.\,3.). To accomplish this, the frequency must be set above the hysteretic region and slowly reduced to the desired value. In the absence of turbulence, the behavior of the MEMS under free decay is determined by solving Eq.\,\ref{eq:eom} for $g_0 = 0$ \cite{Polunin2016}. Here the response of the MEMS is given by $x(t) = A(t)\cos(\omega_0t + \phi(t))$ with

\begin{equation}
A(t) = \frac{A_0e^{-\Gamma_1t}}{\sqrt{1+\frac14 \frac{\Gamma_2}{\Gamma_1}A_0^2(1-e^{-2\Gamma_1t})}}
\label{eq:ringdown_envelope}
\end{equation}
and
\begin{equation}
\dot{\phi}(t) = \Pi A^2(t),
\label{eq:ringdown_phase}
\end{equation}
where $A_0$ is the initial displacement amplitude. The device parameters can then be determined by fits to either the frequency response or the free decay.

In presenting Eqs.\,\ref{eq:frequency_amplitude}\,\--\,\ref{eq:ringdown_phase}, we stated that they only hold true in the absence of turbulence. This is because the functional form of the damping and frequency shifts on the device due to the vorticity is not known {\it a priori} and is not included in Eq.\,\ref{eq:eom}. The response of the device presented above describes the intrinsic behavior of the device, and any deviations may be attributed to the effect of turbulence in the fluid.
 
\section{Results}
\subsection{Frequency Domain}
\begin{figure}
	\centering
	\includegraphics[width=0.96\linewidth]{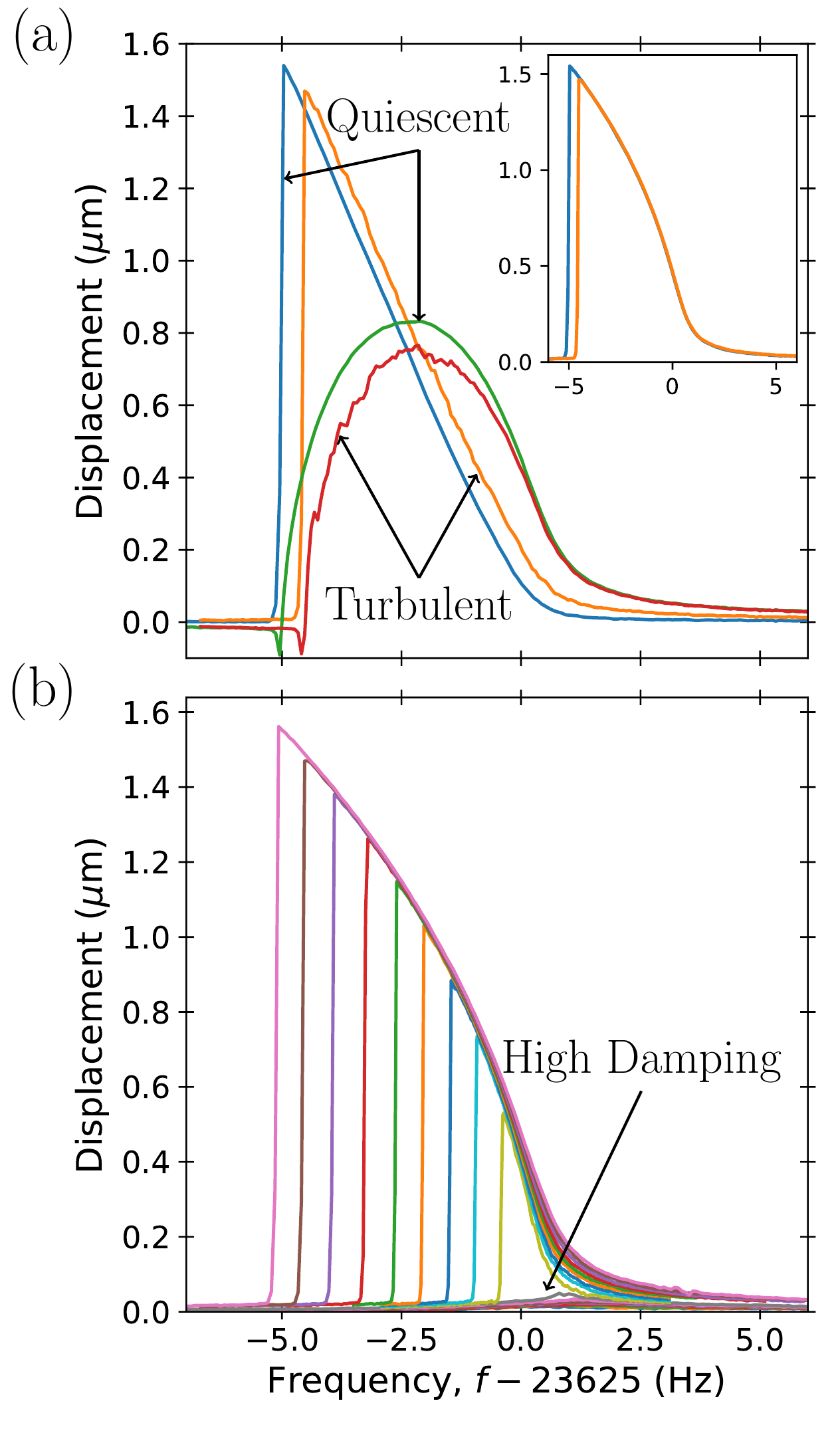}
	\caption{Frequency response of the MEMS in the presence of turbulence generated by the TF measured at 14\,mK. (a) Two downward frequency sweeps of the MEMS with (Turbulent) and without (Quiescent) the TF generating turbulence. The main figure shows the quadrature components of the signal and the inset shows the amplitude. In the turbulent state the peak amplitude is smaller due to increased damping. Excess noise in the phase also appears caused by fluctuations of the damping \cite{csbthesis,Barquist2019}. (b) A set of frequency sweeps made with excitations between 100\,--\,420\,mV$_p$ in steps of 20\,mV$_p$while the TF was generating turbulence ($v_{TF} = 126\,$mm\,s$^{-1}$). For the lowest several excitations, the corresponding sweeps (labeled ``High Damping") experience significant damping, which can be seen by the relatively small amplitudes. This behavior can be seen more clearly in Fig.\,\ref{fig:5}.}
	\label{fig:4}
\end{figure}

The effect of turbulence on the device can be clearly seen by comparing two sweeps made with the same driving force. Figure \ref{fig:4}(a) shows two down sweeps of the MEMS shear mode, both made with 400\,mV$_{p}$ excitation. One sweep was made in the presence of turbulence, labeled ``Turbulent", and the other in its absence, labeled ``Quiescent". For the turbulent sweep, the velocity of the TF was 126\,mm\,s$^{-1}$. There are several features that distinguish the turbulent sweep from the quiescent sweep. The turbulent sweep transition between bi-stable states of the Duffing oscillator (the big jump around 23621\,Hz) occurs at a higher frequency and lower amplitude, and also has its phase shifted relative to the other sweep. Extra noise is also observed in the quadrature channels, but not in the amplitude, which can be interpreted as phase noise. Measurements of the phase noise spectra and discussion of the origin of the noise are to be presented in a forthcoming publication. Because the peak occurs when the driving force is equal to the damping force, and because both sweeps were performed with the same excitation, the lower peak of the turbulent sweep indicates increased damping due to the presence of turbulence. The overall shift in phase and the increase in frequency of the bi-stable transition are consequences of the peak occurring at a lower amplitude. 

\begin{figure}
	\centering
	\includegraphics[width=\linewidth]{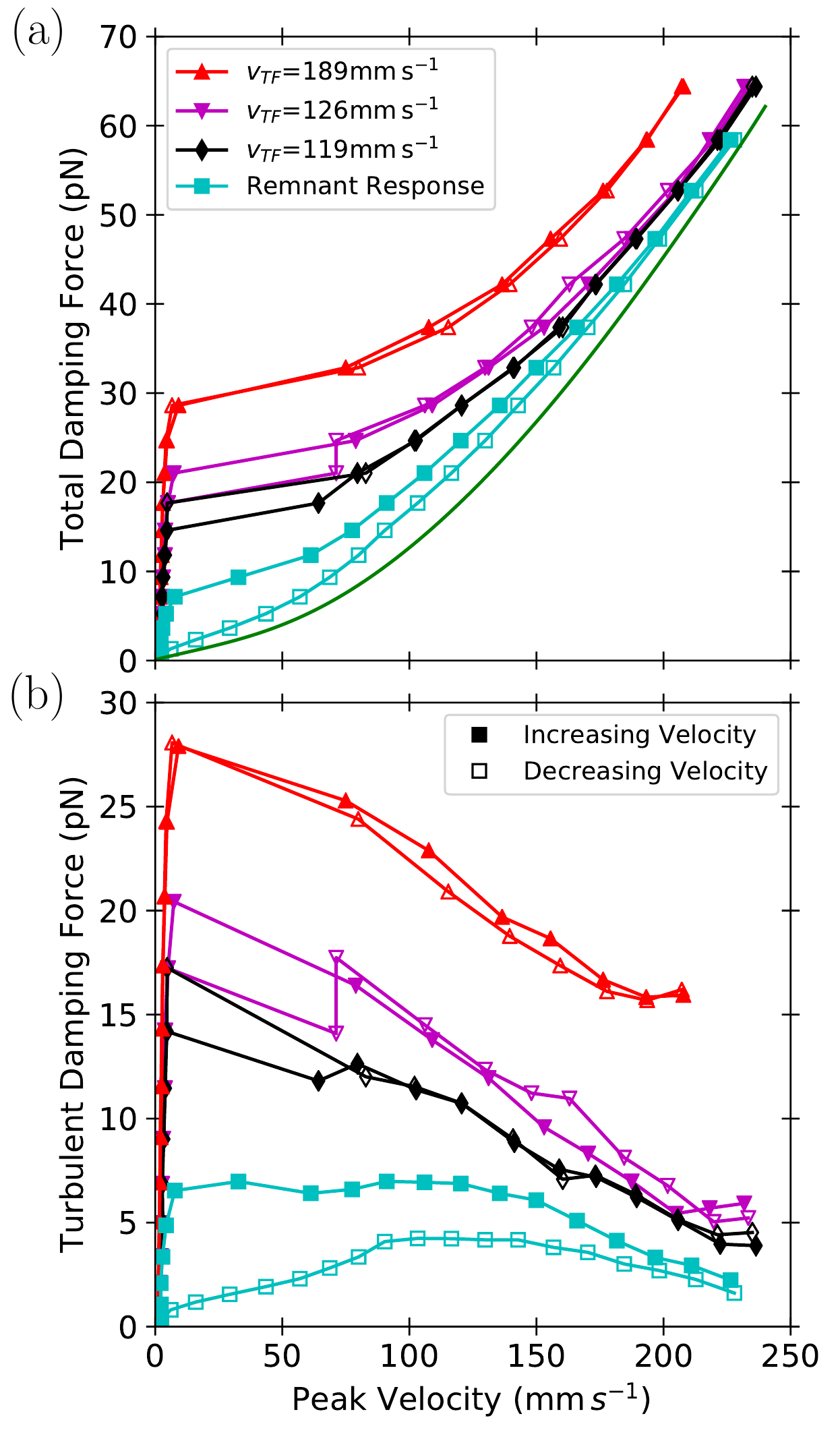}
	\caption{Damping force on the MEMS in the presence of turbulence generated by the TF measured by performing frequency sweeps. Also shown is the damping from remnant vortices, which is measured while the TF is at rest.  (a) The total damping force experienced by the MEMS as a function of velocity. The colored and open symbols represent measurements made for increasing and decreasing MEMS velocity, respectively. (b) The damping force experienced by the MEMS due to turbulence for various TF velocities. The turbulent damping force is calculated by subtracting the intrinsic damping of the device, which is shown as the solid curve in (a). The intrinsic damping was measured in vacuum at 6\,mK.}
	\label{fig:5}
\end{figure}

Figure\,\ref{fig:4}(b) shows a series of frequency sweeps made while the TF was generating turbulence at 126\,mm\,s$^{-1}$. The frequency sweeps were made with excitations from 100--420\,mV$_{p}$ with 20\,mV$_{p}$ steps. It can be seen that for the lowest few excitations the damping is significantly higher compared to the other sweeps (labeled ``High Damping"), and on this scale the signal is indistinguishable from the noise floor. To characterize the velocity dependence of the damping, we record the velocity of the MEMS at the peak for various driving forces. Figure\,\ref{fig:5} shows the result of this measurement. Figure\,\ref{fig:5}(b) depicts the data shown in Fig.\,\ref{fig:5}(a) with the vacuum damping (solid green curve in Fig.\,\ref{fig:5}(a)) subtracted. Therefore, the damping presented in Fig.\,\ref{fig:5}(b) may be identified as the damping from the presence of vortices and turbulence. A preliminary version of the results shown in Fig.\,\ref{fig:4} and Fig.\,\ref{fig:5} were discussed previously in Refs. \cite{Barquist2019,Barquist2019a} 

In Fig.\,\ref{fig:5}, alongside the measurements made in TF generated turbulence, are measurements of the response of the device due to remnant vortices pinned to the surface, which are shown as open and closed squares. These vortices became pinned to the surface after some turbulent event, such as driving the TF or cooling through the superfluid transition \cite{Awschalom1984}. They remain attached to the surface after the turbulence has dissipated. These measurements were made directly after cooling to base temperature through the superfluid transition, so that many remnant vortices were present, and the MEMS was highly damped. They were also made in the absence of turbulence generated by the TF. The measurements were first made for increasing velocity, then decreasing velocity. A large hysteresis between these measurements can be seen and is identified with the removal of some vortices pinned to the device. We identify the change in damping around $v \simeq 5\,$mm\,s$^{-1}$ with the onset of vortex removal. To observe the hysteresis again, more remnant vortices must be generated. Otherwise, upon increasing the velocity again, the damping closely follows the lower curve. We term the process of vortex removal as {\it annealing}. While the MEMS may be annealed by use of the shear mode, as seen in Fig.\,\ref{fig:5}, it may be annealed to a greater degree by driving other modes of the device. The effects of remnant vortices and the annealing process are discussed further in Refs. \cite{csbthesis,Barquist2019,Barquist2019a}.

In the presence of turbulence, hysteresis is no longer observed. This is consistent with the interpretation that pinned vortices are being removed when the velocity of the device exceeds $v \simeq 5$\,mm\,s$^{-1}$. In the turbulent flow, vortices are continually colliding with the MEMS and becoming pinned; any vortices removed by driving the MEMS are quickly replenished. For velocities below 5\,mm\,s$^{-1}$, the MEMS experiences significant drag. This corresponds to the ``High Damping" regime referenced in Fig.\,\ref{fig:4}. Upon exceeding this critical velocity, the damping of the MEMS is reduced, whereupon the velocity of the MEMS jumps to $\simeq70\,$mm\,s$^{-1}$. Similar to the remnant vortex response, in the high velocity regime the turbulent damping force is reduced as the velocity increases.

To understand the high damping regime it is helpful to consider how much extra vortex line length would need to be generated to account for the observed damping. For this, we consider how much energy is dissipated during each cycle of oscillation. We simplify the argument by assuming that the force and velocity are sinusoidal, and that they are exactly in phase. This last assumption is justified because we are only considering the damping on resonance. In this case, the energy dissipated per cycle is $E = F_0v_0/2f$, where $F_0$ and $v_0$ are the amplitudes of the force and velocity, respectively. For $F_0 = 10^{-11}$\,N and $v_0 = 10$\,mm\,s$^{-1}$, the energy dissipated each cycle is $E =2.1\times 10^{-18}$\,J. The linear energy density of a vortex is $E/l = (\rho \kappa^2/4\pi) \ln(\ell/a_0)$ \cite{Donnelly2005}, where $\rho$ is the density of helium, $\ell$ is the inter vortex spacing around the device, and $a_0 \simeq 10^{-10}$\,m is the size of the vortex core. For our device, a reasonable guess for the vortex spacing is in the range $\ell \sim 1 - 100\,\mu$m. This then yields a linear energy density $E/l = 1.0 - 1.4 \times 10^{-18}$\,J\,$\mu$m$^{-1}$, which corresponds to $\sim 2 \,\mu$m/cycle of total increased length.

\begin{figure}
	\includegraphics[width=1\linewidth]{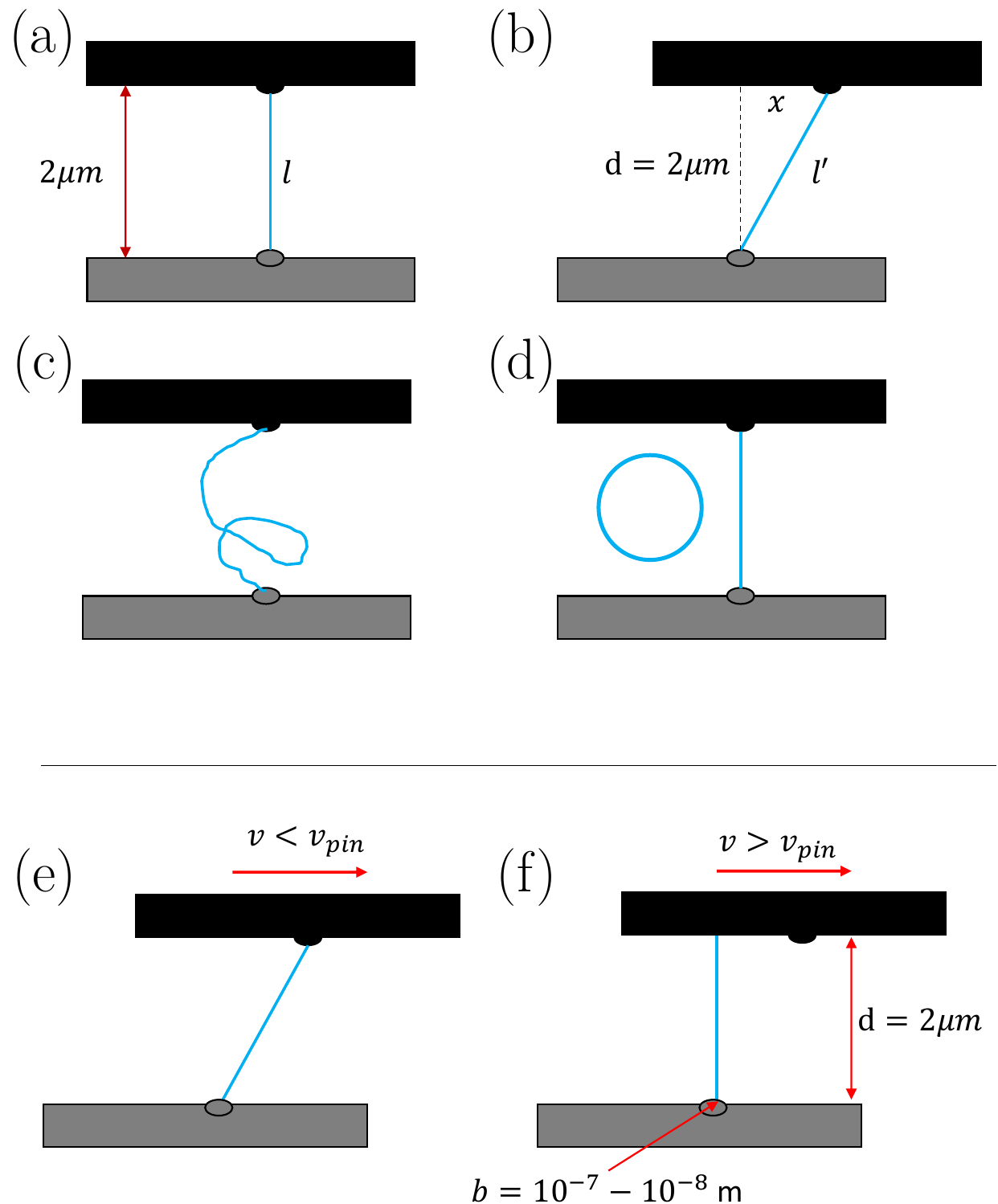}
	\caption{A cartoon of the interaction between the MEMS and quantized vortices. (a)\,--\,(d) A mechanism for damping the MEMS. (a) A quantized vortex is pinned between the device and the substrate. (b) Because of the pinning, the vortex is elongated when the device is displaced. (c) The extra length after displacement is accumulated. Some number of cycles later the vortex line is long enough to reconnect with itself. (d) The reconnection ejects a vortex ring, which removes length from the pinned vortex, and the process begins again. (e)\,--\,(f) A mechanism for the decoupling of the motion of the MEMS from the motion of vortices. (e) At velocities lower than the velocity required to depin a vortex, $v_{pin}$, the motion of the MEMS is coupled to the motion of the vortex. (f) When the MEMS exceeds $v_{pin}$ the vortex is free to slide along the surface and the motion is no longer coupled. A distribution in depinning velocities would cause the damping to be gradually reduced as the velocity is increased, as seen in Fig.\,\ref{fig:4}(b).}
	\label{fig:6}
\end{figure}

We can estimate the number of vortices pinned to the device by constructing a simple model for how the vortices are pinned, see Fig.\,\ref{fig:6}(a)\,--\,(d). The movable portion of the MEMS is suspended $d =2\,\mu$m above a substrate. Consider a straight vortex bridging the moving plate and the substrate such that its length is $d$. If the plate is displaced horizontally by a distance $x$, the vortex length is increased by $\delta l \simeq x^2/2d$. For a complete cycle of oscillation the increased length is twice that amount. The displacement amplitude of the MEMS near the critical velocity is roughly $x \simeq 0.1\,\mu$m. This yields an increase of length per vortex of $\delta l \simeq 5\times 10^{-3}\,\mu$m, which suggests there are about 400 vortices pinned to the device in the high damping regime. This excess vortex length would then be carried away from the device by vortex rings that are created when the length of an individual vortex is large enough to intersect itself and cause a reconnection event \cite{Koplik1993,schwarz1985three}. It is not possible to know the exact distribution of vortices on the device; however, they are most likely concentrated around the perimeter of the device. Including the electrodes, the perimeter of the MEMS is quite large. For 400 vortices the inter-vortex distance is $\ell \sim 1-10\,\mu$m, consistent with our initial guess above.

This simple model is consistent with the linear scaling of the force with velocity in the high damping regime: the energy lost per cycle is proportional to the velocity squared, $E \propto v^2$, and the energy transferred to the vortices is proportional to the increased length which is proportional to the velocity squared, $E \propto \delta l\propto x^2 \propto v^2$. To effectively transfer energy from the device to the vortex line, the frequency of the device should be matched to the frequency of a standing mode of the vortex \cite{vinen2004nucleation}. If the frequency of the device is too low, the vortex line will respond adiabatically and will be in its instantaneous equilibrium position determined by the flow around the device. In this limit, there is no accumulation of excess length through one period of motion. If the frequency of the device is much larger than the standing mode, the coupling of motion is greatly reduced. The frequency of standing wave modes for a quantized vortex is given by
\begin{equation}
f(k) = \frac{1}{2\pi} \frac{\kappa k^2}{4\pi}\ln\left(\frac{1}{ka_0}\right),
\end{equation}  
where $k = n\pi/l$ is the wave number \cite{Donnelly2005}. Taking the range of lengths, $l = 2.0$--$2.2\,\mu$m, we find that the fundamental frequency of the standing mode is in the range 27.3--22.9\,kHz, which is consistent with the frequency of the SH mode (23.6\,kHz), indicating that the device can efficiently transfer energy to the vortices.

Another possible mechanism for the damping of the MEMS that might be proposed is the removal of energy through a Kelvin wave cascade. The initial stages of this process are similar to what is described above, {\it i.e.,} the motion of the device causes motion on the vortices pinned to the device. Except, instead of the energy being carried away from the device as vortex rings, it is ferried to smaller scales of wave motion on the attached vortices through the Kelvin wave cascade until the vortex can efficiently radiate phonons. However, this process predicts that the power dissipated should scale as the tenth power of the displacement amplitude, $A^{10}$ \cite{Eltsov2020}, which is clearly not observed, so this process can not fully explain the damping observed by the device.

Because the TF is situated above the MEMS, it is reasonable to expect that some vortices will become pinned to the top of the moving plate ({\it i.e.} the side in contact with the bulk). However, these vortices should not significantly contribute to the damping because it is unlikely that their lengths correspond to a Kelvin wave resonance at the frequency of the device. It may be expected that vortices pinned to the top of the device could cause the MEMS to transition to turbulent flow in a manner similar to the TF, but this is not observed. Because of the geometry of the device, the backflow around the device is minimal, and it is the backflow which provides the superflow necessary for the transition to turbulence. For these reasons, we do not believe that vortices pinned to the top of the MEMS significantly contribute to the observed results.

The reduction of damping with increasing velocity, observed above the critical velocity, can be understood as the depinning of vortices from pinning sites on the surface. Figures\,\ref{fig:6}(e)\,--\,(f) illustrate this process. When the velocity of the plate is below the depinning velocity, $v_{pin}$, the vortex remains attached to the pinning site. Because the vortex is pinned, its length is increased as the plate is displaced, which leads to damping as discussed above. When the velocity exceeds $v_{pin}$ the vortex is no longer pinned to a specific place on the surface and is free to move relative to the surface. Because of this, the motion of the vortex is no longer coupled to the motion of the plate and no longer contributes to the damping. The depinning velocity for a vortex pinned between two parallel plates was considered by Schwarz \cite{schwarz1985three}, and  was found to be
\begin{equation}
v_{pin} = \frac{\kappa}{2\pi d}\ln\left(\frac{b}{a_0}\right).
\label{eq:depinning_velocity}
\end{equation}
Here $d = 2\,\mu$m is the MEMS gap size and $b = 10^{-7}$--$10^{-8}$\,m is the size of the pinning site, measured by atomic force microscopy \cite{Gonzalez2013}. However, the argument presented in Ref.\,\cite{schwarz1985three} is for uniform superflow between the plates, which is not the case for the MEMS. A vortex pinned between the moving plate and the substrate will experience some velocity gradient as the plate is displaced. Despite this, we use Eq.\,\ref{eq:depinning_velocity} for an order of magnitude comparison. For our MEMS, $v_{pin} \simeq 50\,$mm\,s$^{-1}$, which is close to the region where the damping is observed to decrease, $70$--200\,mm\,s$^{-1}$. The observed velocity dependence of the damping may be due to a distribution in the size of pinning sites. The surface of the MEMS is rough on the scale of 100\,nm, and there is a distribution of bump sizes that make up this roughness. A detailed discussion of the surface characteristics of these MEMS devices is provided in Refs.\,\cite{Gonzalez2013,gonzalezthesis}. Because of the distribution in the bumps there is a distribution in the depinning velocities. As the velocity is increased, more vortices decouple and the damping is reduced. However, the velocity depends only on the logarithm of the bump size, and this is unlikely able to explain this behavior over the whole velocity range. A complete explanation of this phenomena is not possible because there is still much unknown about how the MEMS interacts with the turbulence from the TF. We currently do not know how the MEMS captures and removes vortices. It is also unknown how the flow field induced by the turbulent vortices around the device affects its motion.
\subsection{Time Domain}

We further investigate the effects of turbulence on the MEMS motion by studying the properties of the free decay of the MEMS in the turbulent and quiescent regime. In the quiescent regime we investigate the effect of remnant vortices. For all measurements, the MEMS was tuned to $f_d =23,621.72$\,Hz and was driven at 400\,mV$_p$. The lock-in was referenced at $f_d$, and the time constant was chosen such that $1/\tau >\omega_d -\omega(t)$ for all time. Here, $\omega(t) = \omega_0 + \Pi A^2(t)$ (Eq.\,\ref{eq:ringdown_phase}) depends on the amplitude and changes during the ringdown measurement due to the nonlinear restoring force. Note that the measurements do not begin on resonance. This is done because the noise due to turbulence will readily cause the MEMS to transition from the high amplitude state into the low amplitude state if the frequency is tuned too close to the transition frequency \cite{Stambaugh2006}.

\begin{figure}
	\includegraphics[width=1\linewidth]{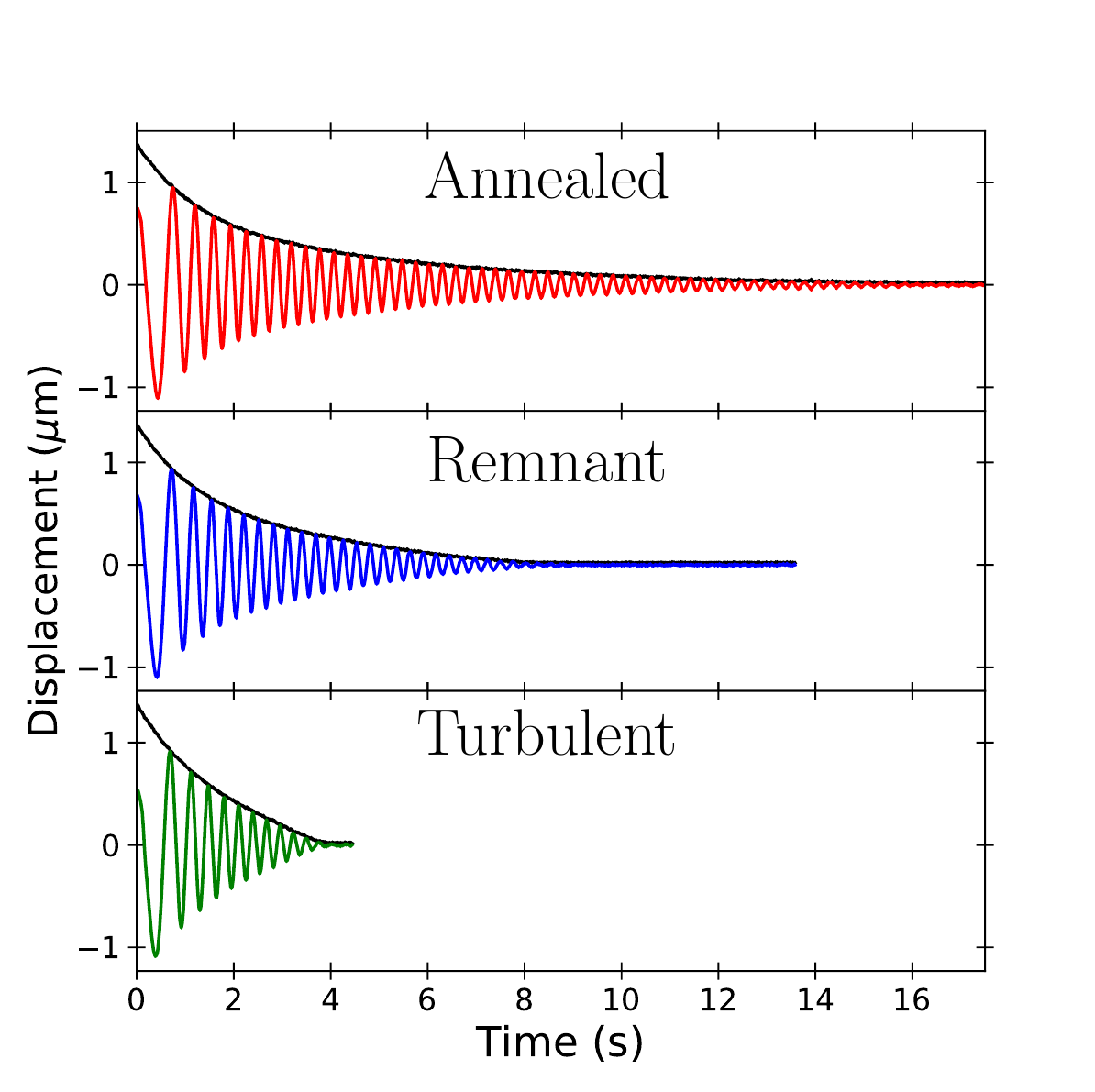}
	\caption{From top to bottom, the response of the MEMS in the quiescent state after being annealed (see text), in the quiescent state before annealing, and with turbulence. The solid black curves represent the amplitude of the device, $A(t)$, while the oscillating curves represent the component of the signal in-phase with the original driving signal, $A(t)\cos[(\omega(t)-\omega_d)t]$ (see Eq.\,\ref{eq:ringdown_phase}). At large amplitude, $\omega(t) \simeq \omega_d$, and the oscillation is slow. As the amplitude is decreased $\dot{\phi}$  tends towards zero and $\omega(t)$ tends toward $\omega_0$. At low amplitude, the signal then oscillates with frequency $|\omega_0-\omega_d|$. The out-of-phase signal is also recorded, but is not plotted for clarity. The envelopes of these ringdowns are shown in Fig.\,\ref{fig:8}b in log-linear scale, where the departure from a pure exponential decay can be easily noticed.}
	\label{fig:7}
\end{figure}

\begin{figure*}
	\centering
	\includegraphics[width=1\linewidth]{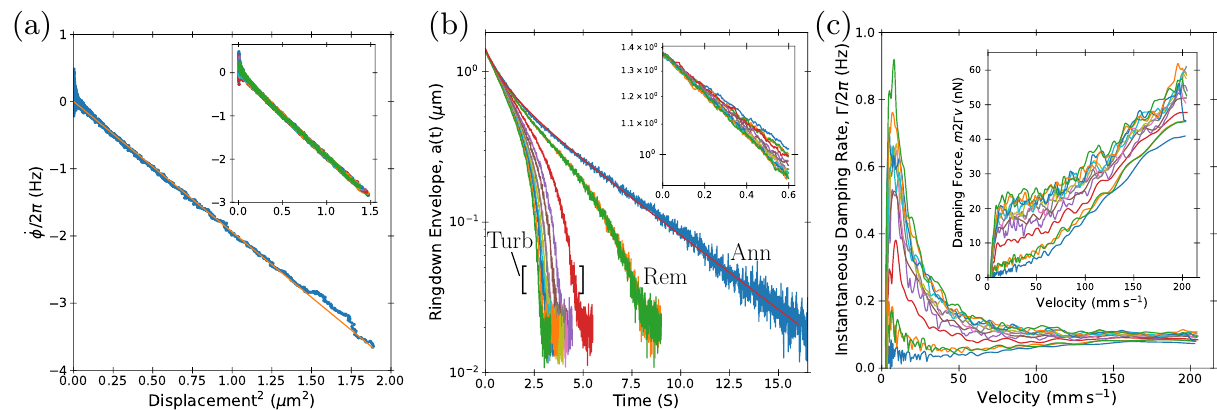}
	\caption{Ringdown response of the MEMS for several different TF velocities between 126-183\,mm\,s$^{-1}$. (a) Amplitude dependent frequency shift calculated from the time dependent phase (see Eq.\,\ref{eq:phi_calculation}). The main figure shows the response in the quiescent state after annealing, and the solid line is a fit to Eq.\,\ref{eq:ringdown_phase} with $\Pi/2\pi = 2.0\,$Hz$\,\mu$m$^{-2}$. The inset shows the frequency response for all different TF velocities, including the response shown in the main figure. Within the precision of the measurement, the presence of turbulence does not alter the linear or nonlinear restoring force. (b) Ringdown envelopes in log-linear scale for various TF velocities. The measurements are grouped into three categories: ``Turb" for measurements made in the turbulent state, ``Rem" for measurements made in the quiescent state before annealing, and ``Ann" for measurements made in the quiescent state after annealing. The solid red line is a fit of the ``Ann" data to Eq.\,\ref{eq:ringdown_envelope} with $\Gamma_1/2\pi = 0.037\,$Hz and $\Gamma_2/2\pi = 0.18\,$Hz\,$\mu$m$^{-2}$. The inset shows the short time behavior at high velocity before the nonlinearities significantly affect the response of the device. (c) The velocity dependence of the instantaneous damping rate, $\Gamma$, calculated from Eq.\,\ref{eq:differential_damping_rate}. $\Gamma$ can be understood as the instantaneous slope of an envelope shown in (b). Inset: Damping force experienced by the MEMS calculated as $m2\Gamma v$, where $m$ is the mass of the MEMS, $v$ is the velocity, and $\Gamma$ is the value displayed in the main figure.}
	\label{fig:8}
\end{figure*}

From top to bottom, Fig.\,\ref{fig:7} shows the ringdown response of the MEMS in the quiescent regime after annealing, in the quiescent regime before annealing, and in the turbulent regime. For each different measurement shown, ten individual decays were recorded and averaged. The black curve is the amplitude of motion, $A(t)$, and the oscillating curve is the in-phase component of the motion, $A(t)\cos[(\omega(t)-\omega_d)t]$. The out-of-phase component is also collected, but it is not plotted for clarity. The frequency of oscillation at any given time is $|f(t) - f_d|$. This shift in frequency is due to demodulation occurring within the lock-in. Relative to the motion in the quiescent annealed state, the motion in the presence of remnant vortices and turbulence is more damped, which is consistent with the frequency domain measurements. The decay in all three cases begins roughly the same, and only deviates as the amplitude is decreased. In the turbulent state, the decay clearly deviates from an exponential time dependence at lower amplitudes. Several more measurements of the free decay of the MEMS were made for various TF velocities in the range of 126\,--\,183\,mm\,s$^{-1}$. Again, for each TF velocity ten decays were measured and averaged. The results of these measurements are shown in Fig.\,\ref{fig:8}.

Visually, it is not obvious from Fig.\,\ref{fig:7} what effect the turbulence has on the nonlinear frequency shift. To see how the turbulence affects the frequency shift we first need to calculate the frequency of oscillation as the amplitude decays. To do this, we first calculate the phase of the oscillator as a function of time from the measured quadrature components as 
\begin{equation}
\phi(t) = \tan^{-1}\left(\frac{A(t)\sin[(\omega(t)-\omega_0)t]}{A(t)\cos[(\omega(t)-\omega_0)t]}\right).
\label{eq:phi_calculation}
\end{equation}
To obtain the total accumulated phase, $\pi$ is added every time the argument of Eq.\,\ref{eq:phi_calculation} changed signs from $-$ to $+$.  The frequency shift was then calculated by numerically differentiating the phase with respect to time. At low amplitude, the noise would cause the sign of the phase to fluctuate and spoil the process described above. A moving average including the 40 nearest points was performed to reduce the noise.

The amplitude dependent frequency shift for the MEMS measured in the quiescent annealed state is shown in Fig.\,\ref{fig:8}(a). The solid line is a fit to the data using Eq.\,\ref{eq:ringdown_phase}. Around $A = 1.3\,\mu$m the frequency shift deviates from the fit. This anomaly is due to the ringdown beginning off resonance. The fit yields a nonlinear frequency pulling of $\Pi/2\pi = -2.0\,$Hz\,$\mu$m$^{-2}$. In the inset of Fig.\,\ref{fig:8}(a), the nonlinear frequency shifts of the MEMS in both the turbulent and quiescent state are shown. Within the precision of the measurement, all of the data lie on top of each other. This implies that the presence of turbulence does not significantly affect the resonance frequency or the nonlinear frequency shift of the MEMS.


The ringdown envelopes for all measurements are shown in Fig.\,\ref{fig:8}(b) in log-linear scale. When plotted this way, the slope at any point is the instantaneous damping rate, and a straight line in the plot corresponds to a pure exponential decay. The envelope labeled ``Ann" was made in the quiescent state after annealing. The solid red line is a fit to the data using Eq.\,\ref{eq:ringdown_envelope}. From the fit we obtain $\Gamma_1/2\pi = 0.037$\,Hz and $\Gamma_2/2\pi = 0.18\,$Hz\,$\mu$m$^{-2}$. Because the normal fluid is absent and the remnant vortices have been removed, the measured linear and nonliner damping may be attributed to intrinsic processes polysilicon. Two measurements were made in the quiescent state directly after turbulence was present, and are labeled ``Rem". These measurements were made before annealing, so the device is still influenced by remnant vortices. The envelopes labeled ``Turb" were made while the TF was continuously generating turbulence, with velocities in the range 126--183\,mm\,s$^{-1}$. Even in the absence of turbulence, the remnant envelope differs significantly from the annealed envelope. Directly after initiating the ringdown, the damping rate begins to decrease. However, as the velocity is reduced further, the damping rate begins to increase for the remnant case, while it remains constant after annealing. The increased damping rate at low amplitude is more pronounced for the measurements made in turbulence. These observations are consistent with our previous measurements: there is large damping at low velocities, and at large velocities the damping is reduced.

The decay of the MEMS in the presence of vorticity cannot be described by Eq.\,\ref{eq:ringdown_envelope}, as the functional form of the damping is unknown. However, we can extract the local damping rate, $\Gamma$, at a given time by computing the slope of $\ln(A(t))$ at that time, that is
\begin{equation}
\Gamma = - \frac{d\ln(A(t)/A_0)}{dt}.
\label{eq:differential_damping_rate}
\end{equation}
The damping rates extracted this way are shown in Fig.\,\ref{fig:8}(c) as a function of the velocity of the resonator. Because of the noise, the data was first smoothed using a local weighted regression and the numerical derivative was averaged over the 100 nearest points. The damping rates are peaked at low velocity, and as the velocity increases the damping falls off like $v^{-1}$ until it begins to plateau around 100\,mm\,s$^{-1}$. The inset of Fig.\,\ref{fig:8}(c) shows the damping force, $m2\Gamma v$, calculated from the data shown in the main figure. This can be directly compared with Fig.\,\ref{fig:5}(a), where it can be seen that the same broad features are displayed. The critical velocity observed in Fig.\,\ref{fig:5}(a) can be understood as the velocity at which the damping rate becomes inversely proportional to $v$, {\it i.e. }$\Gamma \propto v^{-1}$. The large jump in velocity above the critical velocity seen in Fig.\,\ref{fig:5}, can also be understood from the velocity dependence of $\Gamma$. When $\Gamma \propto v^{-1}$, the damping force experienced by the MEMS, $F \propto \Gamma v$, is constant and independent of velocity. Therefore, any incremental increase in the force will cause the velocity to grow until some new process alters the velocity dependence of $\Gamma$ causing the damping force to once again equal the driving force.

For the data presented in Fig.\,\ref{fig:8}, it is important to remember that the data are collected for decreasing velocity. That is, the damping starts off small and is increased as the velocity is reduced. This increase in damping must come from an increased coupling of the device motion to the motion of vortices surrounding the device. This increased coupling can be understood as the same process described in Fig.\,\ref{fig:6}(e)--(f). Also, as the MEMS slows down it is more likely for a vortex line impingent on the device to pin to the surface upon collision. The increased likelihood of vortex capture is due to the increased number of stable pinning sites at lower velocities.

\begin{figure}
	\includegraphics[width=\linewidth]{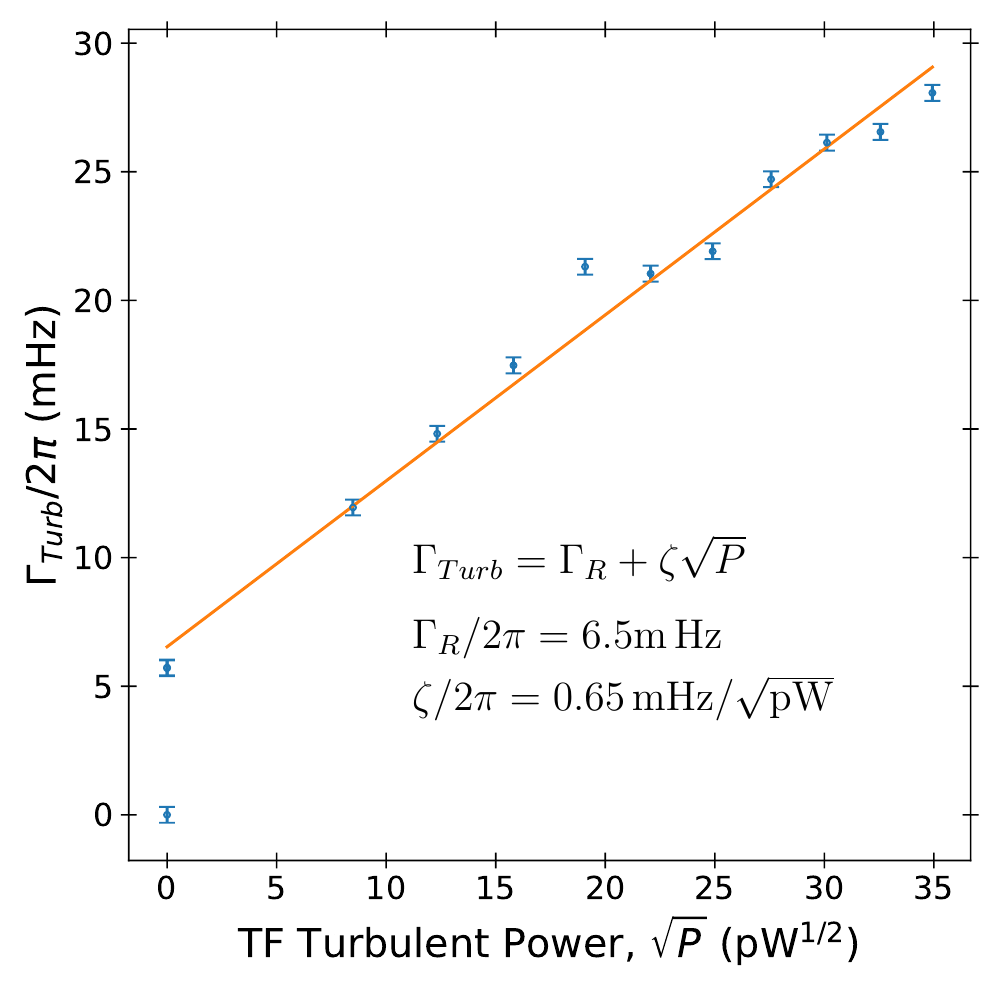}
	\caption{The increase of the MEMS damping rate at high velocity as a function of the square root of the TF power input into turbulence. The damping rate is calculated by fitting a line to the data shown in the inset of Fig.\,\ref{fig:8}(b) for times less than 0.6\,s. The increase in the damping rate due to turbulence, $\Gamma_{Turb}$ is calculated by subtracting the damping rate measured without turbulence or remnant vortices, $\Gamma_{A}$, {\it i.e.} $\Gamma_{Turb} = \Gamma -\Gamma_{A}$.}
	\label{fig:9}
\end{figure}

For short times after the beginning of the ringdown, the damping can be considered to be linear, which is demonstrated in the inset of Fig.\,\ref{fig:8}(b). By fitting the data in the first 0.6\,s to a line, the damping rate, $\Gamma$ can be extracted. The increase in damping rate arising from the turbulence, $\Gamma_{Turb}$, can be determined by subtracting the damping rate of the device in the annealed state, $\Gamma_{A}$, {\it i.e.} $\Gamma_{Turb} = \Gamma - \Gamma_{A}$. It is found that $\Gamma_{Turb}$ grows in proportion the the square root of the turbulent power. This is demonstrated in Fig.\,\ref{fig:9}, where $\Gamma_{Turb}$ is plotted against the square root of the turbulent power. The solid line is a fit to the data, and the fitting function with the fit values are shown in the figure. The turbulent power is calculated from the excess force experienced by the TF when it is driving turbulence, $F_T = F-\gamma v$, where $\gamma$ is determined by fitting in the potential flow regime, see Fig,\,\ref{fig:2}. The power input to turbulent flow is then $P = F_{T}v$. 

The dependence of $\Gamma_{Turb}$ on turbulent power may be understood by assuming that we are in the ultra-quantum regime of turbulence \cite{Walmsley2014,Baggaley2012,Skrbek2012}. This regime is distinct from the quasi-classical regime discussed in the introduction, and is characterized by a lack of large scale flow. Here, the turbulent vortex tangle is random and lacks any significant polarization. In other words, most of the turbulent energy is contained at the scale of the inter vortex spacing, $\ell$. With $\ell$ being the characteristic length scale of the ultra-quantum turbulence, it may be shown by dimensional analysis, that the decay rate of the vortex line density, $L$, is proportional to $\kappa L^2$ \cite{1957a}. To determine the steady state value of $L$ we must add a term to account for the creation of line length coming from the TF. Because the energy of a vortex line is proportional to its length, the rate of increase of line length should be proportional to the power $P$. Including this we obtain
\begin{equation}
\frac{dL}{dt} = \chi_PP - \chi_2 \kappa L^2,
\end{equation} 
where $\chi_P$ is a proportionality constant, which can be understood at the inverse of the average linear energy density of a vortex in the turbulent flow. The steady state value of $L$ is found when $dL/dt = 0$, yielding
\begin{equation}
L = \sqrt{\frac{\chi_PP}{\kappa \chi_2}}.
\end{equation}
It is natural to assume that the damping on the device is proportional to the number of vortices interacting with the device, and in turn proportional to $L$. Assuming this, we arrive at $\Gamma_{Turb} \propto L \propto \sqrt{P}$. Unfortunately, we are unable to make a quantitative calculation of $L$, because we lack a precise theory for how $L$ corresponds to $\Gamma$.  However, by making a similar measurement for a greater range of TF velocities, we may be able to observe the crossover from the ultra-quantum to quasi-classical regime of turbulent flows. In the fully developed quasi-classical regime $dL/dt \propto -L^{3/2}$, which yields $\Gamma_{Turb} \propto P^{2/3}$. 

The offset of $\Gamma$ at $P=0$, $\Gamma_R$, is likely due to a semi-permanent background of remnant vortices. The data point at (0 pW, 6.5\,mHz) corresponds to the damping of both remnant ringdown measurements, which were made at different times. One was made after driving the TF at 125\,mm\,s$^{-1}$ and the other after driving it at 183\,mm\,s$^{-1}$. Although they were made at separate times, their ringdown envelopes overlap almost perfectly. This indicates that there are some long lived remnant vortices that are not removed by simply driving the shear mode to high velocity. The extra damping of these vortices is then present for all of the ringdown measurements made in turbulence, which is seen as a constant shift of $\Gamma_{Turb}$. If there really is a semi-permanent background of remnant vortices, then why do the remnant ringdowns differ from the annealed ringdown (see Fig.\,\ref{fig:8}(b))? Before making the annealed measurement, the device was thoroughly annealed using a combination of different resonance modes of the device, and was annealed for a longer time. The extra care in annealing seems to be responsible for the removal of these semi-permanent vortices.

\section{Conclusion}
We have presented measurements of our MEMS device in the presence of turbulence generated by a secondary structure (tuning fork) in the ultra-low temperature regime (14\,mK) and demonstrated that our device is uniquely sensitive to turbulence. The uniqueness of this device is its ability to continuously measure the turbulent flow. Until now, all other oscillators measured in superfluid $^4$He begin to generate their own turbulence immediately after being exposed to vorticity in the fluid. While this has enabled many rich experiments \cite{Yano2010,nago2011time,kubo2013time,nago2013vortex,oda2014observations,Yano2019}, it precludes the use of these devices to continuously sense quantum turbulence.
 
To demonstrate the sensitivity of the MEMS, we have presented measurements of the device in both the frequency and time domain. It was observed that below a critical velocity of about 5\,\,mm\,s$^{-1}$, the damping of the MEMS is greatly enhanced relative to its intrinsic damping. Above the critical velocity, the damping is greatly reduced. From the time domain measurements, it was observed that this critical velocity corresponds to a change in the velocity dependence of the damping, with the damping rate changing inversely proportional to the velocity, $\Gamma \propto v^{-1}$. To explain the damping at low velocities and the change in damping above the critical velocity we propose a model of vortices pinned between the substrate and the moving part of the MEMS. The model accounts for the reduction in damping above the critical velocity by supposing there is a distribution in depinning velocities, and that when a vortex become depinned its motion is decoupled from the motion of the device. However, from the frequency domain measurements of the device under the influence of remnant vortices, it is clear that complete removal of some fraction of the vortices is also happening above the critical velocity. Through the time domain measurements, it was also found that high velocity damping of the MEMS scales in proportion to $\sqrt{P}$. We interpret this result as an indication that the MEMS is sensing the average vortex line density, $L$, which scales as $\sqrt{P}$ in the ultra quantum regime.

While there is still much to learn about the interaction of vortices with the MEMS device before more quantitative statements can be made, it is clear that there is a wealth of information that can be extracted from these devices.


%



\begin{acknowledgments}
We thank Mark Dykman and Wei Guo for thoughtful discussions about the interpretations of our results. This work is supported by the National Science Foundation through the grant DMR-1708818.
\end{acknowledgments}

\bibliographystyle{./apsrev4-2}
\bibliography{csb-prx-2020}

\end{document}